\documentclass{article}
\usepackage{amssymb,bm}

\newcommand{\sakujo}[1]{}

\newcommand{\ABS}[1]{#1}
\newcommand{\EIG}{\lambda}
\newcommand{\HALL}{\alpha}
\newcommand{\HEL}{\sigma}
\newcommand{\POL}{s}

\newcommand{\Elsasser}{Els\"asser}

\newcommand{\LieBrace}[3]{{#1\{}#2,#3{#1\}}}
\newcommand{\LieBracket}[3]{{#1[}#2,#3{#1]}}

\newcommand{\Braket}[3]{{#1<}#2{#1|}#3{#1>}}

\newcommand{\GEV}[2]{\vec{\bm{#1}}#2}
\newcommand{\GEC}[2]{\widehat{#1}#2}
\newcommand{\GEM}[2]{\tilde{#1}#2}

\newcommand{\CHM}[1]{\hat{#1}}
\newcommand{\CHW}[2]{{\bm{#1}}#2}
\newcommand{\CHC}[2]{\widehat{#1}#2}

\newcommand{\EIGEN}[2]{\lambda^{{#2}}(\widetilde{#1})}

\newcommand{\tripleHMHD}[4]{{#1(}\!{#1(}#2{#1|}\!{#1|}#3{#1|}\!{#1|}#4{#1)}\!{#1)}}
\newcommand{\tripleEuler}[4]{{#1(}#2{#1|}#3{#1|}#4{#1)}}

\newcommand{\dd}[2]{\frac{\partial #2}{\partial #1}}
\newcommand{\sdd}[2]{\frac{{\rm{d}} #2}{{\rm{d}} #1}}

\newcommand{\DivFree}[2]{{#1(}#2{#1)}_{S}}

\newcommand{\CalcNote}[1]{}%
\renewcommand{\CalcNote}[1]{*>#1<*}%
\renewcommand{\sakujo}[1]{/*#1*/}%

\newcommand{\q}{x}

\title{
Differential-geometrical approach to
the dynamics of dissipationless incompressible Hall magnetohydrodynamics:
I.
Lagrangian mechanics 
on semidirect product of two volume preserving diffeomorphisms
and conservation laws
}
\author{Keisuke ARAKI\\
\it
Okayama University of Science,\\
\it
1-1 Ridai-cho, Okayama, 700-0005 JAPAN\\
\it
E-mail: araki@are.ous.ac.jp
}

\begin{document}

\maketitle

\begin{abstract}
The dynamics of 
a dissipationless incompressible Hall magnetohydrodynamic (HMHD) medium 
is formulated
using Lagrangian mechanics 
on a semidirect product of two volume preserving diffeomorphism groups.
In the case of $\mathbb{T}^3$ or $E^3$,
the generalized {\Elsasser} variables introduced by Galtier 
(S. Galtier 2006 J. Plasma Phys. \textbf{72} 721-69)
yield
remarkably simple expressions of basic formulas and equations
such as
the structure constants of Lie algebra,
the equation of motion,
and
the conservation laws.
\textit{Four} constants of motion,
where three of the four are independent,
are naturally derived from
the generalized {\Elsasser} variables representation of 
the equation of motion for the HMHD system:
total plasma energy, magnetic helicity, hybrid helicity,
and the modified cross helicity.

\end{abstract}

\section{Introduction}

About a half century ago, 
when Arnold reviewed his study on the dynamical systems on Lie groups
and related hydrodynamic topics
in a unified form, 
he treated three materials in somewhat intertwining manner
(Arnold 1966), these were:
(1) Lagrangian mechanics on Lie groups and 
derivation of the equation of motion
including exposition of its relation to Noether's theory, 
(2) Hamiltonian mechanics on Lie groups and 
the consideration of stability problems around stationary solutions,
and 
(3) reinterpretation of the equation of motion
as a geodesic equation 
under an appropriate Affine connection on a Lie group
and its application to predictablity problems 
in terms of the Riemannian curvature induced by the connection.
Since then,
these three topics have evolved separately.
Today,
the Hamiltonian mechanics on a wide varieties of Lie groups
are summarized 
in terms of the Lie-Poisson structure 
of the cotangent bundle of Lie groups
(Marsden and Ratiu 1994, Morrison 1998 for comprehensive reviews)
and 
the Euler-Poincare structure
has been established 
as its Lagrangian mechanical counterpart
(Holm Marsden and Ratiu 1998 for a comprehensive review)
and
a number of dynamical systems have been reformulated 
according to these approaches.

In the present study,
we treat Hall magnetohydrodynamic (HMHD) systems
from a Lagrangian mechanical viewpoint.
HMHD systems are regarded as one of the simplest models of a plasma
that includes the so-called two fluid effect,
i.e.,
the effect arising from 
the difference between the ion and electron current field
(Lighthill 1960).
In HMHD formulation, Ohm's law is approximated as follows:
\begin{eqnarray}
 \bm{E} + \frac{1}{c}\bm{u}\times\bm{B}
 =
 \frac{1}{qnc} \bm{j}\times\bm{B}
,
\end{eqnarray}
where $\bm{E}$, $c$, $\bm{u}$, $\bm{B}$, $q$, $n$, $\bm{j}$ 
are electric field, the speed of light, 
mean velocity of the plasma, magnetic field,
electric charge, number density of ions, and
current field, respectively.
The Hall effect term on the right hand side
reflects
the influence of the finite size of the ion skin depth. 
HMHD systems have been investigated intensively
in the context of both space and fusion plasmas.

The effects of the Hall term 
on the properties of magnetohydrodynamic (MHD) turbulence 
have been investigated from various viewpoints
including
the closure approach in the weak/wave turbulence framework 
(Galtier 2006),
evaluation of turbulent energy transfer and dynamo action
using direct numerical simulation (DNS) data
(Mininni et al. 2007),
and
coherent structure formation by DNS
(Miura and Araki 2014).

From a mathematical perspective,
HMHD systems have gradually attracted the attention of researchers 
for
such topics as
the conservation laws of Lagrangian invariant form
({Shivamoggi 2009}, {Araki 2009}),
existence of a solution to initial value problem
({Acheritogaray et al. 2011}, {Chae et al. 2014}).
In terms of the analytical mechanical studies of HMHD systems,
since Holm's pioneering work in this field ({Holm 1987}),
Hamiltonian approaches have predominated
({Sahraoui et al. 2003},
{Hirota et al. 2006},
{Yoshida and Hameiri 2013}).
Keramidas Charidakos et al. 
did however employ the Lagrangian approach to
derive the HMHD momentum equation and Ohm's law
using the first variation of action
constructed from a general two-fluid plasma Lagrangian
(Keramidas Charidakos et al. 2014).
To the author's knowledge,
little research has focused on
the Lagrangian mechanical approach associated with
the geodesic/the Riemannian curvature formulation
which latter originates from Arnold's third formulation,
although
much literature exists regarding standard MHD system
({Zeitlin and Kambe 1993}, {Hattori Y 1994}).
In this study, 
we examine HMHD system from Arnold's first formulation
in the 1966 article.
We will present
a mathematical model that derives the HMHD equations 
as Euler-Lagrange equations for variational problems
and
examine its configuration space, described using a Lie group.

In the present study,
we focus on the three-dimensional case
and
do not treat the more general $N$-dimensional case 
for the sake of mathematical simplicity.

\vspace{1em}

This paper is organized as follows:
in section 2,
the foundations of Lagrangian mechanics is discussed 
and
the equation of motion derived from Hamilton's principle.
In section 3,
we discuss the Lie algebraic structure of 
the function space of the velocity and magnetic fields.
The equation of motion, 
expanded by the generalized {\Elsasser} variables (GEV),
is reviewed from a differential geometrical viewpoint
in Section \ref{generalized Elsasser variables representation}.
Section \ref{constants of motion from Z-representation viewpoint}
is devoted to the derivation of the constants of motion
from the GEV derived equation of motion.
The summary is presented in section \ref{sec:discussion}.

\section{Configuration space and Lagrangian mechanics of HMHD systems}

Although a number of continuum systems 
have been investigated 
using semidirect product groups, 
most of them are ``passive'' to the second element of the product,
i.e.,
\begin{eqnarray}
\label{action of GxV}
 ( g_{1}, v_{1}) \circ ( g_{2}, v_{2})
 &=&
 ( g_{1} \circ g_{2},  v_1 + B(g_1) v_2 )
,
\end{eqnarray}
where 
$g_{1}$, $g_{2} \in G$, $v_{1}$, $v_2 \in V$,
and
$B$: $G \times V \to V$ is 
an appropriate representation of $G$ on a vector space $V$
(Marsden and Ratiu 1994, Arnold and Khesin 1998).
An exception is the case treated by Vizman (2001),
in which the product is given by 
\begin{eqnarray}
\label{action of GxH}
 ( g_{1}, h_{1}) \circ ( g_{2}, h_{2})
 &=&
 ( g_{1} \circ g_{2},  h_{1} \circ B(g_1) h_2 )
,
\end{eqnarray}
where 
$h_{1}$, $h_{2}$ are the elements of a Lie group $H$, 
$B$: $G \times H \to H$ is group homomorphism.
As we will demonstrate,
the dynamics of an HMHD system is founded on this group action
and induced Lie algebra
so that
we will follow the work of Vizman in the present study.

\vspace{1em}

Let $M$ be 
a container of an HMHD fluid that is
a three-dimensional differentiable manifold
with coordinate system $({\q}^{i})$.
The basic variables of an HMHD system in this study 
are a pair of volume preserving diffeomorphism maps on $M$
denoted by
$
 (\vec{X},\vec{Y})
$
where
$\vec{X}$, $\vec{Y}\in S$Diff($M$).
\footnote{%
In this paper,
we use arrow symbol on top of 
a multifunctional character of mathematical quantities.
%
For example,
diffeomorphism (a triplet of functions) 
is expressed by $\vec{X}=(X^{1},X^{2},X^{3})$,
and 
a pair of vector fields by $\GEV{Z}{}=(\bm{u},\bm{b})$.
Boldface letters are used to denote vector fields on $M$.
}%

The first element
$
 \vec{X}=\vec{X}(\vec{\q},t)=\big(X^{i}(\vec{\q},t)\big)
$ 
is a triplet of functions on $M$.
The value $\vec{X}(\vec{a},t)$ 
physically implies the position of a fluid particle at time $t$,
which was initially located at $\vec{a}\in M$.
The map $\vec{X}$ is related to the velocity field
in the Eulerian specification
$\bm{u}(t)=u^{i}(t)\dd{{\q}^{i}}{}\in \mathfrak{X}_{\Sigma}(M)$
by the formula
$$
  \left.\dd{\tau}{X^{i}(\vec{a},\tau)}\right|_{\tau=t}
  \left(\dd{{\q}^{i}}{}\right)_{\vec{\q}=\vec{X}(\vec{a},t)}
  =
  u^{i}\left(\vec{X}(\vec{a},t),t\right)
  \left(\dd{{\q}^{i}}{}\right)_{\vec{\q}=\vec{X}(\vec{a},t)}
,
$$
where $\mathfrak{X}_{\Sigma}(M)$ is
the Lie algebra of $S$Diff($M$),
i.e.,
the function space of divergence-free vector fields
equipped with Lie bracket $[\bm{a},\bm{b}]$.
We denote the space of $\vec{X}$ by $G$ hereafter.

The second element
$
 \vec{Y}= \vec{Y}(\vec{a},s;t)=
 \big(Y^{i}(\vec{a},s;t)\big)
$
is a ``stream function'' of the current field
where 
$s$ and $t$ are the current line parameter and time, respectively.
The map $\vec{Y}$ is related to the current field
in the Eulerian specification
$\bm{j}(t)=j^{i}(t)\dd{{\q}^{i}}{}\in \mathfrak{X}_{\Sigma}(M)$
by the formula
$$
  \dd{s}{Y^{i}(\vec{a},s;t)}
  \left(\dd{{\q}^{i}}{}\right)_{\vec{\q}=\vec{Y}(\vec{a},s;t)}
  =
  -\HALL \, j^{i}\left(\vec{Y}(\vec{a},s;t),t\right)
  \left(\dd{{\q}^{i}}{}\right)_{\vec{\q}=\vec{Y}(\vec{a},s;t)}
,
$$
where $\HALL$ is a parameter 
that determines the relative strength of the Hall term.
We denote the space of $\vec{Y}$ by $H$ hereafter.

Group operation of a diffeomorphism, $S$Diff($M$),
is given by the composition of the function triplets
and we denote it by
$
 \vec{X}_{1} \circ \vec{X}_{2} = \vec{X}_{1}(\vec{X}_{2})
.
$
By defining the group operation 
for the pairs of basic variables ($\vec{X}$,$\vec{Y}$)
by the function compositions
$$
 \left(\vec{X}_{1},\vec{Y}_{1}\right) \circ \left(\vec{X}_{2},\vec{Y}_{2}\right)
 =
 \left(
  \vec{X}_{1}(\vec{X}_{2})
 ,
  \vec{Y}_{1}\big(\vec{X}_{1}\big(\vec{Y}_{2}(\vec{X}_{1}^{-1})\big)\big)
 \right)
,
$$
we obtain the semidirect product of the diffeomorphisms $G \ltimes H$.

Substituting an exponential map approximation
for each variable
$
 \vec{X}_{k}(\tau)=\exp(\tau\bm{u}_{k})
$ 
and
$
 \vec{Y}_{k}(\tau;t)=\exp(-\tau\HALL\bm{j}_{k}(t))
,
$
using Hausdorff's formula 
$\exp(\bm{a})$ $\exp(\bm{b})$ $=$ 
$\exp(\bm{a}+\bm{b}+\frac12[\bm{a},\bm{b}]+\cdots)$
and 
taking $O(\tau^2)$ terms of function compositions, 
we obtain the following Lie bracket 
on $\mathfrak{g} \ltimes \mathfrak{h}$,
which defines the Lie algebra of $G \ltimes H$:
\begin{eqnarray}
&&
 \LieBracket{\Big}{\GEV{V}{_{1}}}{\GEV{V}{_{2}}}
 =
 \bigg(
  \nabla \times ( \bm{u}_{1} \times \bm{u}_{2} )
 ,
 \nonumber\\&&\hspace{2em}
  - \HALL
  \Big(
     \nabla \times ( \bm{u}_{1} \times \bm{j}_{2} )
   + \nabla \times ( \bm{j}_{1} \times \bm{u}_{2} )
   - \HALL \nabla \times( \bm{j}_{1} \times \bm{j}_{2} )
  \Big)
 \bigg)
,
\label{eq:Lie bracket in V-space}
\end{eqnarray}
where
${\GEV{V}{_{k}}}:=\big(\bm{u}_{k},-\HALL\bm{j}_{k}\big)$ 
is
a tangent vector of one parameter subgroup
$\gamma_{k}(\tau):=$ ( $\vec{X}_{k}(\tau)$, $\vec{Y}_{k}(\tau;t)$ ),
i.e.,
a small path on the configuration space $G \ltimes H$,
and
$\nabla\times$ is the curl operator 
in standard vector analysis notation.
%
%
Since $M$ is three-dimensional and 
the vector fields considered here are divergence free,
the Lie bracket on 
$\mathfrak{X}_{\Sigma}(M)$
is given by $[\bm{a},\bm{b}]=\nabla\times(\bm{a}\times\bm{b})$.
Note that,
if the semidirect product operation (\ref{action of GxV}) is used, 
the current fields product term in (\ref{eq:Lie bracket in V-space}) 
does not appear in the Lie bracket,
and thus, 
the evolution equation of standard one-fluid MHD is obtained 
(Hattori 1994).

Let $\gamma(t)$ ($t\in[0,1]$) be a path on $G \ltimes H$
on which
a sufficiently short interval between the two points 
$\gamma(t)$ and $\gamma(t+\tau)$ 
is well approximated by
$\big($
 $\exp\big(\tau\bm{u}(t)\big)$, 
 $\exp\big(-\tau\HALL\bm{j}(t)\big)$
$\big)$.
To consider the variational problem along this path,
we exert a perturbation on the path
$
 \gamma(t;\delta)=
 \left(
   \exp\big(\delta\,\bm{\xi}(t)\big) , 
   \exp\big(-\delta\HALL\bm{\eta}(t)\big)
 \right)
 \circ\gamma(t;0)
,
$
where
$\delta$ is a small parameter,
$
 \gamma(t;0)=\gamma(t)
,
$
and
$\bm{\xi}$, $\bm{\eta}$
are the ``displacement'' fields.
Let
$
 \GEV{V}{(t;\delta)}
 =
 \GEV{V}{(t)}
 +
 {\delta}\,\GEV{v}{(t)}
 +
 o(\delta)
$
be the tangent vector along the perturbed path
where 
$\GEV{v}{(t)}$
is the perturbation part of the tangent vector.
For the case of the dynamical systems on Lie groups,
the perturbation part $\GEV{v}{(t)}$
is known to satisfy Lin's constraint
(Cendra and Marsden 1987)
\renewcommand{\upsilon}{\nu}
\begin{eqnarray}
 \GEV{v}{}
 =
 \dot{\GEV{\upsilon}{}}
 +
 \LieBracket{\big}{
  \GEV{\upsilon}{}
 }{
  \GEV{V}{}
}
,
\end{eqnarray}
where
$
 \GEV{\upsilon}{}
 :=
 \big( \bm{\xi}(t), -\HALL\,\bm{\eta}(t) \big)
.
$

To define the Riemannian metric on $G \ltimes H$,
we introduce the inverse of the curl operation,
which corresponds physically to the calculation of 
the magnetic field $\bm{b}$ induced by a current field $\bm{j}$.
Hereafter, $(\nabla\times)^{-1}$
denotes the operation to obtain
the vector potential $\bm{b}$
from a divergence-free field $\bm{j}$
that satisfies
\begin{eqnarray}
  \bm{j}=\nabla\times\bm{b},\ 
  \nabla\cdot\bm{b}=0,\ 
  \bm{b}^{(H)}=\bm{0}
,
\end{eqnarray}
where
$\nabla\cdot$ is the divergence operator
and
$\bm{b}^{(H)}$ is harmonic function component of
Hodge-Kodaira decomposition of $\bm{b}$ (Yoshida and Giga 1990).

Here we define Riemannian metric on $G\ltimes H$ by
\begin{eqnarray}
&&\hspace{-2em}
 \Braket{\Big}{\GEV{V}{_{1}}}{\GEV{V}{_{2}}}_{(\vec{X},\vec{Y})}
 :=
 \int_{\vec{a}\in M}\Big(
  \bm{u}_{1} \cdot \bm{u}_{2}
 \Big)_{\vec{X}(\vec{a},t)}
 {\rm{d}}^3\vec{X}(\vec{a},t)
 \nonumber\\&&
 +
 \int_{\vec{b}\in M}\Big(
  \big((\nabla\times)^{-1}\bm{j}_{1}\big)
  \cdot
  \big((\nabla\times)^{-1}\bm{j}_{2}\big)
 \Big)_{\vec{Y}(\vec{b},s;t)}
 {\rm{d}}^3{\vec{Y}(\vec{b},s;t)}
,
\end{eqnarray}
where ${(\vec{X},\vec{Y})} \in G\ltimes H$
and
$
 {\rm{d}}^3{\vec{X}(\vec{a},t)}
,
$
$
 {\rm{d}}^3{\vec{Y}(\vec{b},s;t)}
$
are the advected volume elements initially located
at $\vec{a}$ and $\vec{b}$, respectively.
Since the maps $\vec{X}$ and $\vec{Y}$ are volume preserving:
$
 {\rm{d}}^3{\vec{\q}}=
 {\rm{d}}^3{\vec{X}(\vec{\q},t)}=
 {\rm{d}}^3{\vec{Y}(\vec{\q},s;t)}
,
$
the value of the Riemannian metric is invariant
against the right-translation.
%
%
%
We define the Lagrangian 
\begin{eqnarray}
 L=
 \frac12
 \Braket{\big}{\GEV{V}{}}{\GEV{V}{}}_{(\vec{X},\vec{Y})}
\label{eq:lagrangian V-sp}
\end{eqnarray}
to be the total plasma energy.
Note that
this Lagrangian is the same as that used in Hattori (2011)
to derive 
the equation of motion of an incompressible standard MHD fluid.
The first variation of action associated with this Lagrangian becomes
\begin{eqnarray}
&&\hspace{-3em}
 \sdd{\delta}{ S}
 =
 \int_{0}^{1}
  \Braket{\big}{
   \GEV{V}{}
  }{
   \GEV{v}{}
  }_{(\vec{X},\vec{Y})}
 {\rm{d}}t
 =
 \int_{0}^{1}
  \Braket{\big}{
   \GEV{V}{}
  }{
   \GEV{v}{}
  }_{(e,e)}
 {\rm{d}}t
 =
 \int_{0}^{1}
  \Braket{\big}{
   \GEV{V}{}
  }{
   \dot{\GEV{\upsilon}{}}
   +
   \LieBracket{\big}{
    \GEV{\upsilon}{}
   }{
    \GEV{V}{}
   }
  }_{(e,e)}
 {\rm{d}}t
\nonumber\\&&\hspace{-3em}
 =
 \int_{0}^{1} {\rm{d}}t
 \int_{\vec{\q}\in M} {\rm{d}}^3\vec{\q}\bigg[
  \bm{u} \cdot \Big(
   \dot{\bm{\xi}}
   +
   \nabla \times ( \bm{\xi} \times \bm{u} )
  \Big)
  \nonumber\\&&
  +
  \bm{b} \cdot \DivFree{\big}{
   \dot{\bm{\beta}}
   + ( \bm{\xi} \times \bm{j} )
   + ( \bm{\eta} \times \bm{u} )
   - \HALL ( \bm{\eta} \times \bm{j} )
  }
 \bigg]
 \label{eq:variational calculation process}
,
\end{eqnarray}
where 
$e$ is the identity map,
$\bm{b}:=(\nabla\times)^{-1}\bm{j}$,
$\bm{\beta}:=(\nabla\times)^{-1}\bm{\eta}$
and hereafter
$\DivFree{}{*}$ is 
the projection onto the solenoidal component of a vector field.
Integration by parts with fixed end conditions yields
the following Euler-Lagrange equation:
\begin{eqnarray}
&&
 \dot{\bm{u}}
 =
 \DivFree{\big}{
  \bm{u} \times
  (\nabla \times \bm{u})
  +
  \bm{j} \times \bm{b}
 }
,
\label{eq of mot u}
\\&&
 \dot{\bm{b}}
 =
 \nabla \times ( \bm{u} \times \bm{b} )
 -
 \HALL
 \nabla \times ( \bm{j} \times \bm{b} )
.
\label{eq of mot b}
\end{eqnarray}
Thus,
the evolution equation of 
an incompressible ideal Hall magnetohydrodynamic medium
is derived as 
a dynamical system on a semidirect product of two diffeomorphism groups.
Note that
in the limit $\HALL\to0$,
we obtain the standard MHD model for incompressible plasmas.
For the purely hydrodynamic case ($\bm{j}=\bm{b}=\bm{0}$),
the Euler equation for an incompressible fluid is obtained.

As is obvious from our formulation,
by setting the Hall strength parameter to $\HALL=0$
in (\ref{eq:Lie bracket in V-space}),
one simply obtains the Euler equation.
The second element of $\GEV{V}{}$-variables 
should be taken as such a vector field
that
needs to be divided by $\HALL$ to obtain $O(1)$ vector field.
Thus,
the standard MHD system should be considered 
as a kind of singular limit of present formulation.
Analytical mechanical approaches to the HMHD systems
seem to raise such small parameter problem
when its relation to the standard MHD limit is considered.
For example,
in the Hamiltonian mechanics approach,
one of the natural choices of vector variables 
is the pair of
the total ion momentum density
$
 \bm{M}=\rho \bm{v} + R^{-1} a \rho \bm{A}
$
and
the magnetic vector potential $\bm{A}$,
where $R/a=\HALL$ in our notation (Holm 1987).
In the limit $\HALL\to0$,
these two variables come close to each other 
$\bm{M}\approx R^{-1} a \rho \bm{A}$,
and 
manipulation of the small difference $\bm{v}=\bm{M}/\rho - R^{-1} a \bm{A}$
is needed to capture the ion flow.

\section{Lie algebra induced on the space of ($\bm{u}$, $\bm{b}$)}

For practical purposes,
it is convenient to use the magnetic field $\bm{b}$
instead of the current field $\bm{j}$.
Here,
we introduce 
a pair of velocity and magnetic fields
by
$
 \GEV{Z}{} = \big( \bm{u}, \bm{b} \big)
.
$
%
%
From
the variational calculation process 
(\ref{eq:variational calculation process}), 
the bilinear operation on the $\GEV{Z}{}$-variable space
defined by the following formula
is expected to work as a commutator of certain Lie algebra:
\begin{eqnarray}
&&
 \LieBrace{\Big}{\GEV{Z}{_{1}}}{\GEV{Z}{_{2}}}
 =
 \bigg(
  \nabla \times ( \bm{u}_{1} \times \bm{u}_{2} )
 ,
 \nonumber\\&&\hspace{2em}
   \DivFree{\Big}{
       \bm{u}_{1} \times (\nabla\times\bm{b}_{2})
     + (\nabla\times\bm{b}_{1}) \times \bm{u}_{2}
     - \HALL (\nabla\times\bm{b}_{1}) \times (\nabla\times\bm{b}_{2})
   }
 \bigg)
,
\label{eq:Lie bracket in Z-space}
\end{eqnarray}
where
$
 \GEV{Z}{_{k}}:=(\bm{u}_{k},\bm{b}_{k})
.
$
The definition (\ref{eq:Lie bracket in Z-space}) is skew symmetric.
Introducing the inner product on the $\GEV{Z}{}$-variable space given by
\begin{eqnarray}
\label{eq:Riemannian metric in Z-space}
 \Braket{\Big}{
  \GEV{Z}{_{1}}
 }{
  \GEV{Z}{_{2}}
 }_{Z}
 :=
 \int_{\vec{\q}\in M}\Big(
  \bm{u}_{1} \cdot \bm{u}_{2}
  +
  \bm{b}_{1}
  \cdot
  \bm{b}_{2}
 \Big){\rm{d}}^3\vec{\q}
,
\end{eqnarray}
we see that the value of the Lagrangian (\ref{eq:lagrangian V-sp})
and
the value of the inner product of the commutator and a variable
are preserved under this mapping:
\begin{eqnarray}
&&
  L
  =
  \frac12
  \Braket{\Big}{
    \GEV{V}{}
  }{
    \GEV{V}{}
  }_{(e,e)}
  =
  \frac12
  \Braket{\Big}{
    \GEV{Z}{}
  }{
    \GEV{Z}{}
  }_{Z}
,
\\&&
  \Braket{\Big}{
    \GEV{V}{_{1}}
  }{
    \LieBracket{\big}{
      \GEV{V}{_{2}}
    }{
      \GEV{V}{_{3}}
    }
  }_{(e,e)}
  =
  \Braket{\Big}{
    \GEV{Z}{_{1}}
  }{
    \LieBrace{\big}{
      \GEV{Z}{_{2}}
    }{
      \GEV{Z}{_{3}}
    }
  }_{Z}
,
\end{eqnarray}
where
$
 \GEV{Z}{_{k}}
$
($k=1$, 2, 3)
are induced from
$
 \GEV{V}{_{k}} = \big( \bm{u}_{k}, -\HALL\bm{j}_{k} \big)
$
by 
$\bm{b}_{k}=(\nabla\times)^{-1}\bm{j}_{k}$.
It is straightforward but lengthy to check that
the definition (\ref{eq:Lie bracket in Z-space}) 
satisfies the Jacobi identity.
The commutator defines a Lie algebra on $\GEV{Z}{}$-variable space.

Under this mapping from $\GEV{V}{}$-variable space to 
$\GEV{Z}{}$-variable space,
the variational problem is also formulated on 
$\GEV{Z}{}$-variable space.
The first variation of action is mapped as follows:
\begin{eqnarray}
 \int_{0}^{1}
  \Braket{\big}{
   \GEV{V}{}
  }{
   \dot{\GEV{\upsilon}{}}
   +
   \LieBracket{\big}{
    \GEV{\upsilon}{}
   }{
    \GEV{V}{}
   }
  }_{(e,e)}
 {\rm{d}}t
 =
 \int_{0}^{1}
  \Braket{\big}{
   \GEV{Z}{}
  }{
   \dot{\GEV{\zeta}{}}
   +
   \LieBrace{\big}{
    \GEV{\zeta}{}
   }{
    \GEV{Z}{}
   }
  }_{Z}
 {\rm{d}}t
,
\label{eq:first variation in Z-space}
\end{eqnarray}
where
$
 \GEV{\zeta}{(t)}:=(\bm{\xi}(t),\bm{\beta}(t))
$
is the displacement field in $\GEV{Z}{}$-variable space 
which is mapped from
$
 \GEV{\nu}{(t)}=(\bm{\xi}(t),-\HALL\bm{\eta}(t))
.
$
Thus, 
the Lie algebraic structure on $\GEV{Z}{}$-variable space 
is also expected to
work as a framework of the Lagrangian mechanics
of 
an ideal incompressible HMHD medium.
In the following,
we treat the HMHD system
as a dynamical system on the Lie algebra of $\GEV{Z}{}$-variable space.

\section{Generalized {\Elsasser} variables representation%
\label{generalized Elsasser variables representation}}

In this section,
we demonstrate that
in the case of $M=\mathbb{T}^3$ or $E^3$,
the \textit{generalized {\Elsasser} variables} (GEV)
(Galtier 2006)
yield
remarkably simple expressions of basic formulas or equations
such as
the structure constant of Lie algebra,
equation of motion,
etc.
In this section,
we review Galtier's derivation 
from differential geometrical viewpoint.

Since the velocity and magnetic fields satisfy 
divergence-free conditions
$
  \nabla\cdot\bm{u}
  =
  \nabla\cdot\bm{b}
  =
  0
,
$
they can be expanded in the complex helical wave (CHW) modes as
\renewcommand{\CHM}[1]{\vec{#1},{\HEL_{#1}}}
\begin{eqnarray}
\label{CHW of u b}
  \bm{u}(\vec{x},t)
  =
  \sum_{\CHM{k}}
    \CHC{u}{(\CHM{k};t)}\,
    \CHW{\phi}{(\CHM{k};\vec{x})}
,
\ \ \ 
  \bm{b}(\vec{x},t)
  =
  \sum_{\CHM{k}}
    \CHC{b}{(\CHM{k};t)}\,
    \CHW{\phi}{(\CHM{k};\vec{x})}
,
\end{eqnarray}
where a circumflex denotes a Fourier coefficient
and the function
$
  \CHW{\phi}{(\CHM{k};\vec{x})}
$
is a normalized complex helical wave (CHW).
A CHW is defined by
\begin{eqnarray}
  \CHW{\phi}{(\CHM{k};\vec{x})}
  :=
  2^{-\frac12}
  \bm{h}(\vec{k},\HEL_{k})
  e^{2 \pi i \vec{k}\cdot\vec{x}}
,
\end{eqnarray}
where 
\begin{eqnarray}
  \bm{h}(\vec{k},\HEL_{k}):=\bm{e}_{\theta}(\vec{k})+i\HEL_{k}\bm{e}_{\phi}(\vec{k})
,
\end{eqnarray}
$\HEL_{k}$, $\vec{k}$, $\bm{e}_{\theta}$, $\bm{e}_{\phi}$
are
the helical vector,
helicity ($\HEL_{k}=\pm1$), wavenumber, 
and
the base vector of spherical coordinate system on wavenumber space
in the $\theta$- and $\phi$-directions,
respectively
(Lesieur 1997 Sect.V-5, Waleffe 1992).

To derive the GEV representation of an HMHD system,
we introduce the base functions of $\vec{\bm{Z}}$-variable space,
each of which is given by
\renewcommand{\GEM}[2]{\vec{#1},\HEL_{#1},{#2}\POL_{#1}}%
\renewcommand{\EIGEN}[2]{\lambda_{\HEL_{#1}}^{{#2}\POL_{#1}}(\ABS{#1})}%
\begin{eqnarray}
\label{eigenfunction of L}
  \GEV{\Phi}{(\GEM{k}{};\vec{x})}
  =
  \left(\begin{array}{c}
    \CHW{\phi}{(\CHM{k};\vec{x})}
  \\
    \EIGEN{k}{}\,
    \CHW{\phi}{(\CHM{k};\vec{x})}
  \end{array}\right)
,
\end{eqnarray}
where ${\POL}_{k}=\pm1$,
$\ABS{k}=|\vec{k}|$, 
and 
$\EIGEN{k}{}$ is given by
\begin{eqnarray}
\label{eigenvalue of L}
  \EIGEN{k}{}
  =
  \HEL_{k}\left(
    \POL_{k}\sqrt{(\pi\HALL\ABS{k})^2+1}
    -\pi\HALL\ABS{k}
  \right)
.
\end{eqnarray}
The functions $\GEV{\Phi}{}$
are derived
as eigenfunctions of the linear problem
with a uniform background magnetic field
$\bm{B}_0$:
$
  \partial_t \GEV{Z}{} = \hat{L} \GEV{Z}{}
,
$
where
\begin{eqnarray}
  \hat{L}\GEV{Z}{}
  =
  \left(\begin{array}{c}
    \DivFree{\big}{\bm{j}\times\bm{B}_0}
  \\
    \nabla\times\big((\bm{u}-\HALL\bm{j})\times\bm{B}_0)
  \end{array}\right)
.
\end{eqnarray}
In terms of the complex helical wave mode representation
(\ref{CHW of u b}),
the linear operator $\hat{L}$ can be reduced to
$
  \hat{L} = 2\pi i{B}_0k_{\parallel} \hat{M}
,
$
where
\begin{eqnarray}
  \hat{M}
  =
  \left(\begin{array}{cc}
    O & I \\ I & -\HALL \nabla\times
  \end{array}\right)
  =
  \left(\begin{array}{cc}
    O & I \\ I & -2\pi{\HALL}{\HEL_{k}}\ABS{k}I
  \end{array}\right)
,
\end{eqnarray}
${B}_0=|\bm{B}_0|$,
$k_{\parallel}=\vec{k}\cdot(\bm{B}_0/{B}_0)$
and
$I$ is the identity operator.
The eigenequation of $\hat{M}$ is 
$
 {\EIG}^2 + 2 \pi \HALL \HEL_{k} \ABS{k} \EIG - 1 = 0
$
and
the solutions $\EIGEN{k}{}$ satisfy the following relations:
\begin{eqnarray}
\label{sum of eigenvalues}
&&
  \EIGEN{k}{} + \EIGEN{k}{-}
  = - 2 \pi \HALL \HEL_{k} \ABS{k}
,
\\
\label{product of eigenvalues}
&&
  \EIGEN{k}{}\, \EIGEN{k}{-} = -1
.
\end{eqnarray}
The eigenvalues and eigenfunctions of the operator $\hat{M}$ 
are given by Eqs.(\ref{eigenvalue of L}) and (\ref{eigenfunction of L}),
respectively.
The eigenfunctions physically represent
the ion cyclotron waves ($\POL_{k}=1$ 
for left-hand circular polarization) 
and 
the whistler waves ($\POL_{k}=-1$
for right-hand circular polarization),
respectively
(Galtier 2006, Cramer 2011 \S2.3.1).

Since the eigenvalues (\ref{eigenvalue of L}) are real and non-degenerate,
$
  \GEV{\Phi}{(\GEM{k}{};\vec{x})}
$
are orthogonal:
\begin{eqnarray}
\label{Riemannian metric in Z-coefficient}
  \Braket{\Big}{
    \overline{
      \GEV{\Phi}{(\GEM{k}{};\vec{x})}
    }
  }{
    \GEV{\Phi}{(\GEM{p}{};\vec{x})}
  }_{Z}
  =
  \big(1+(\EIGEN{k}{})^2\big)\,
  \delta^5_{(\GEM{k}{}),(\GEM{p}{})}
,
\end{eqnarray}
where 
an overline denotes a complex conjugate and
$\delta^5$ is a product of Kronecker's deltas.
Note that,
this formula gives component of the Riemannian metric tensor.

\renewcommand{\GEM}[2]{\tilde{#1}#2}
Since the eigenfunctions themselves
do not depend on the intensity of $\bm{B}_{0}$,
they work as orthogonal bases of an HMHD system
even if the uniform background magnetic field is absent.
We introduce here the
expression%
{\footnotemark}:
\begin{eqnarray}
  \GEV{Z}{(\vec{x},t)}=
  \sum_{\GEM{k}{}}
  \GEV{Z}{(\GEM{k}{};\vec{x},t)}
,
\label{GEV expansion of Z}
\end{eqnarray}
where 
the notation $\GEM{k}{}:=$%
$
  \renewcommand{\GEM}[2]{\vec{#1},\HEL_{#1},{#2}\POL_{#1}}
  (\GEM{k}{})
$
is introduced for brevity.
\footnotetext{%
Note that,
since
$
  \EIGEN{k}{} \approx
  \HEL_{k} \POL_{k} - \HEL_{k}(\pi \HALL \ABS{k})
  + \frac12 \HEL_{k} \POL_{k}  (\pi \HALL \ABS{k})^2
  + O(\HALL^4)
,
$
the mode sums
$
 \GEC{Z}{(\vec{k},+,+)}\CHW{\phi}{(\vec{k},+)}
 +
 \GEC{Z}{(\vec{k},-,-)}\CHW{\phi}{(\vec{k},-)}
$, 
$
 \GEC{Z}{(\vec{k},+,-)}\CHW{\phi}{(\vec{k},+)}
 +
 \GEC{Z}{(\vec{k},-,+)}\CHW{\phi}{(\vec{k},-)}
$
converge to 
the {\Elsasser} variables of a standard MHD system
$\bm{z}_{+}=\bm{u}+\bm{b}$, $\bm{z}_{-}=\bm{u}-\bm{b}$
in the limit $\HALL\to0$. 
}%
The component $\GEV{Z}{(\GEM{k}{})}$
and corresponding Fourier coefficient $\GEC{Z}{(\GEM{k}{})}$
are given by
\begin{eqnarray}
&&
  \GEV{Z}{(\GEM{k}{};\vec{x},t)}
  :=
  \GEC{Z}{(\GEM{k}{};t)}
  \GEV{\Phi}{(\GEM{k}{};\vec{x})}
,
\\
  &&
  \GEC{Z}{(\GEM{k}{};t)}
  :=
  \Braket{\Big}{
    \overline{\GEV{\Phi}{(\GEM{k}{};\vec{x})}}
  }{
    \GEV{Z}{(\vec{x},t)}
  }_{Z}
  \Big/
  \Big( 1 + \big(\EIGEN{k}{}\big)^2 \Big)
,
\end{eqnarray}
respectively.
The relation between the complex helical Fourier coefficients
of velocity and magnetic fields
$\CHC{u}{(\CHM{k}{})}$, $\CHC{b}{(\CHM{k}{})}$
and 
the generalized {\Elsasser} variable $\GEC{Z}{(\GEM{k}{})}$ is given by
\renewcommand{\CHM}[1]{\hat{#1}}%
\begin{eqnarray}
\label{relation between Z and (u,b)}
  \GEC{Z}{(\GEM{k}{};t)}
  =
  \frac{
    \CHC{u}{(\CHM{k}{};t)}
    +
    \EIGEN{k}{}\,
    \CHC{b}{(\CHM{k}{};t)}
  }{
    1
    +
    \big(\EIGEN{k}{}\big)^2
  }
,
\end{eqnarray}
where 
the notation 
$
  \CHM{k}{}:=
$%
$
  \renewcommand{\CHM}[1]{\vec{#1},{\HEL_{#1}}}
  (\CHM{k}{})
$
is introduced for brevity.
%
%
The
velocity and magnetic fields are expressed
in terms of $\GEC{Z}{(\GEM{k}{})}$ as follows:
\begin{eqnarray}
&&
\label{u in Z-repr}
  \bm{u}(\vec{x},t)
  =
  \sum_{\GEM{k}{}}
    \GEC{Z}{(\GEM{k}{};t)}\,
    \CHW{\phi}{(\CHM{k};\vec{x})}
,
\\ &&
\label{b in Z-repr}
  \bm{b}(\vec{x},t)
  =
  \sum_{\GEM{k}{}}
    \EIGEN{k}{}\,
    \GEC{Z}{(\GEM{k}{};t)}\,
    \CHW{\phi}{(\CHM{k};\vec{x})}
.
\end{eqnarray}
Substituting the mode expansions 
(\ref{u in Z-repr}) and (\ref{b in Z-repr})
into Eq.(\ref{eq:Lie bracket in Z-space})
and taking the inner product with
$
  \GEV{Z}{(\GEM{k}{};\vec{x})}
,
$
we obtain 
\begin{eqnarray}
&&
  \Braket{\Big}{
    \GEV{Z}{(\GEM{k}{})}
  }{
    \LieBrace{\Big}{
      \GEV{Z}{(\GEM{p}{})}
    }{
      \GEV{Z}{(\GEM{q}{})}
    }
  }_{Z}
  =
  \tripleHMHD{\big}{\GEM{k}{}}{\GEM{p}{}}{\GEM{q}{}}
  \EIGEN{k}{-}\,
  \GEC{Z}{(\GEM{k}{})}\,
  \GEC{Z}{(\GEM{p}{})}\,
  \GEC{Z}{(\GEM{q}{})}
,
\label{<k|{p,q}> in Z-space}
\end{eqnarray}
where the symbol
$\tripleHMHD{}{\GEM{k}{}}{\GEM{p}{}}{\GEM{q}{}}$
is defined by 
the integral of the scalar triple product of the complex helical waves
as follows{\footnotemark}:
\begin{eqnarray}
\label{fac str cont HMHD}
  \tripleHMHD{\big}{\GEM{k}{}}{\GEM{p}{}}{\GEM{q}{}}
  &:=&
  \HALL^{-1}
  \tripleEuler{\big}{\CHM{k}}{\CHM{p}}{\CHM{q}}
  \left(
    \Big( \EIGEN{k}{} \EIGEN{p}{} \EIGEN{q}{} \Big)^2
    -
    1
  \right)
,
\\
\label{def(k|p|q)}
  \tripleEuler{\big}{\CHM{k}}{\CHM{p}}{\CHM{q}}
  &:=&
  \int_{\vec{x}\in M}
    \CHW{\phi}{(\CHM{k};\vec{x})}
    \cdot\Big(
      \CHW{\phi}{(\CHM{p};\vec{x})}
      \times
      \CHW{\phi}{(\CHM{q};\vec{x})}
    \Big)
  {\rm{d}}^3\vec{x}
.
\end{eqnarray}
\footnotetext{%
The value of triple product $\tripleEuler{}{*}{*}{*}$ is given by
\begin{eqnarray}
  \tripleEuler{}{\CHM{k}}{\CHM{p}}{\CHM{q}}
  =
  \frac{
    e^{i\Psi\{\vec{k},\vec{p},\vec{q}\}}|\vec{p}\times\vec{q}|
  }{
    2\sqrt{2}\ABS{k}\ABS{p}\ABS{q}
  }
  (\HEL_{k}\ABS{k}+\HEL_{p}\ABS{p}+\HEL_{q}\ABS{q})
  \delta^3_{\vec{k}+\vec{p}+\vec{q},\vec{0}}
,
\end{eqnarray}
(Waleffe 1992, Galtier 2006),
where the explicit expression of phase factor is
\begin{eqnarray}
  e^{i\Psi\{\vec{k},\vec{p},\vec{q}\}}
  =
  \frac{ikpq}{|\vec{p}\times\vec{q}|^3}
  (\vec{k}\cdot\bm{h}(\vec{p},\HEL_{p}))
  (\vec{p}\cdot\bm{h}(\vec{q},\HEL_{q}))
  (\vec{q}\cdot\bm{h}(\vec{k},\HEL_{k}))
.
\end{eqnarray}
}%
The derivation is given in 
\ref{derivation of the value of Lie bracket in Z-representation}.
The symbols $\tripleHMHD{}{*}{*}{*}$ and $\tripleEuler{}{*}{*}{*}$ 
have the following properties:
(1) they have skew and cyclic symmetry under the permutations of mode indices;
(2) they are non zero if the wavenumbers constitute a closed triangle
$\vec{k}+\vec{p}+\vec{q}=\vec{0}$;
and
(3) they are zero if the wavenumbers are parallel 
$\vec{p}\parallel\vec{q}$.
We obtain $\GEC{Z}{}$-representation of the Lie bracket, 
i.e.,
the structure constants of Lie algebra 
for the generalized {\Elsasser} variables are
as follows:
\begin{eqnarray}
&&
  \LieBrace{\Big}
    {\GEV{Z}{(\GEM{p}{})}}
    {\GEV{Z}{(\GEM{q}{})}}
  =
  \sum_{\HEL_{k},\POL_{k}}
  \tripleHMHD{\big}
    {-\vec{p}-\vec{q},\HEL_{k},\POL_{k}}
    {\GEM{p}{}}{\GEM{q}{}}
  \nonumber\\&&\hspace{3em}\times
  \frac{
    \EIG_{\HEL_{k}}^{-{\POL}_{k}}(|\vec{p}+\vec{q}|)
  }{
    1 + \left(
    \EIG_{\HEL_{k}}^{-{\POL}_{k}}(|\vec{p}+\vec{q}|)
    \right)^2
  }
  \GEC{Z}{(\GEM{p}{})}
  \GEC{Z}{(\GEM{q}{})}\,
  \GEV{\Phi}{(\vec{p}+\vec{q},\HEL_{k},\POL_{k};\vec{x})}
.
\end{eqnarray}
Applying this formula to the expression for the first variation
(\ref{eq:first variation in Z-space}),
we obtain the evolution equation for the generalized {\Elsasser} variables
as follows:
\renewcommand{\GEM}[2]{\tilde{#1}}%
\begin{eqnarray}
  \frac{\partial}{\partial t}
    \overline{\GEC{Z}{(\GEM{k}{};t)}}
  =
  \mathop{
    \sum_{\GEM{p}{}}
    \sum_{\GEM{q}{}}
  }^{\vec{k}+\vec{p}+\vec{q}=\vec{0}}
  \frac{
    \tripleHMHD{}{\GEM{k}{}}{\GEM{p}{}}{\GEM{q}{}}\,
    \EIGEN{q}{-}
  }{
    \left(1+\EIGEN{k}{}^2\right)
  }
  \GEC{Z}{(\GEM{p}{};t)}\,
  \GEC{Z}{(\GEM{q}{};t)}\,
.
\label{basic equation in hatZ}
\end{eqnarray}
The derivation is given in 
\ref{derivation of equation of motion in Z-representation}.

\section{Constants of motion from the $\GEC{Z}{}$-representation perspective
\label{constants of motion from Z-representation viewpoint}}

Using the eigenvalue relation 
(\ref{product of eigenvalues}),
we can rewrite the coefficients that appear in the evolution equation 
(\ref{basic equation in hatZ})
as follows:
\begin{eqnarray}
&&\hspace{-3em}
  \frac{
    \tripleHMHD{}{\GEM{k}{}}{\GEM{p}{}}{\GEM{q}{}}\,
    \EIGEN{q}{-}
  }{
    \left(1+\EIGEN{k}{}^2\right)
  }
 =
 \HALL^{-1}
 \tripleEuler{\big}{\CHM{k}}{\CHM{p}}{\CHM{q}}
 \nonumber\\&&\times
 \frac{
  \EIGEN{k}{} \EIGEN{p}{} \EIGEN{q}{}
  -
  \EIGEN{k}{-} \EIGEN{p}{-} \EIGEN{q}{-}
 }{
  \EIGEN{k}{} - \EIGEN{k}{-}
 }
 \EIGEN{p}{}
.
\label{rewrite the ((k||p||q))}
\end{eqnarray}
%
%
The symmetric properties of this coefficient 
naturally lead to four conservation laws.
Surely,
it is well known that
the incompressible HMHD system has three independent constants of motion,
total plasma energy, magnetic helicity, and hybrid helicity
(Turner 1986).

The first two are 
derived from the skew symmetry of the coefficient
$
  \tripleHMHD{}{\GEM{k}{}}{\GEM{p}{}}{\GEM{q}{}}
.
$
The two identities given by
\begin{eqnarray}
&&
  \mathop{
    \sum_{\GEM{k}{}}
    \sum_{\GEM{p}{}}
    \sum_{\GEM{q}{}}
  }^{\vec{k}+\vec{p}+\vec{q}=\vec{0}}
  \tripleHMHD{}{\GEM{k}{}}{\GEM{p}{}}{\GEM{q}{}}
  \EIGEN{q}{-}\,
  \GEC{Z}{(\GEM{k}{};t)}\,
  \GEC{Z}{(\GEM{p}{};t)}\,
  \GEC{Z}{(\GEM{q}{};t)}\,
  =
  0
,
\\&&
\label{eq:derivation C2 constant}
  \mathop{
    \sum_{\GEM{k}{}}
    \sum_{\GEM{p}{}}
    \sum_{\GEM{q}{}}
  }^{\vec{k}+\vec{p}+\vec{q}=\vec{0}}
  \tripleHMHD{}{\GEM{k}{}}{\GEM{p}{}}{\GEM{q}{}}
  \EIGEN{k}{-}\,
  \EIGEN{q}{-}\,
  \GEC{Z}{(\GEM{k}{};t)}\,
  \GEC{Z}{(\GEM{p}{};t)}\,
  \GEC{Z}{(\GEM{q}{};t)}\,
  =
  0
.
\end{eqnarray}
lead to the following constants of motion
{\footnotemark}:
\begin{eqnarray}
  C_{1}
  &=&
  \sum_{\GEM{k}{}}
    (1+\EIGEN{k}{}^2)
    |\GEC{Z}{(\GEM{k}{};t)}|^2
,
\\
\label{eq:C2 constant}
  C_{2}
  &=&
  \sum_{\GEM{k}{}}
    (\EIGEN{k}{-}-\EIGEN{k}{})
    |\GEC{Z}{(\GEM{k}{};t)}|^2
.
\end{eqnarray}
\footnotetext{%
The problem of the regularity of the $\GEV{Z}{}$-variables that
guarantee the convergence of these identities,
which was discussed for the Euler equation case
by Eyink, Constantin et al.
(Eyink 1994, Constantin, E and Titi 1994),
may remain,
but we assume here appropriate regularity
(cf. Chae Degond Liu 2014).
}%
Note that
the constant $C_{1}$ is twice the total plasma energy
$
 \int(|\bm{u}|^2+|\bm{b}|^2){\rm{d}}^3\vec{x}
.
$
%
%

Derivations of the other two constants of motion 
use the symmetry of the fraction part of
(\ref{rewrite the ((k||p||q))}).
%
%
\newcommand{\EIGENx}[2]{\EIG_{\HEL_{#1}}^{#2}({#1})}%
\newcommand{\GEMx}[2]{\vec{#1},\HEL_{#1},{#2}}%
The evolution equation of 
$
 \EIGENx{k}{+} \GEC{Z}{(\GEMx{k}{+};t)}
 +
 \EIGENx{k}{-} \GEC{Z}{(\GEMx{k}{-};t)}
$
becomes
\begin{eqnarray}
&&\hspace{-2em}
 \frac{\partial}{\partial t}
 \left(
  \EIGENx{k}{+}
  \overline{\GEC{Z}{(\GEMx{k}{+};t)}}
  +
  \EIGENx{k}{-}
  \overline{\GEC{Z}{(\GEMx{k}{-};t)}}
 \right)
 \nonumber\\&&\hspace{-1em}
 =
 \mathop{
  \sum_{\GEM{p}{}} \sum_{\GEM{q}{}}
 }^{\vec{k}+\vec{p}+\vec{q}=\vec{0}}
 \frac{
  \tripleEuler{\big}{\CHM{k}}{\CHM{p}}{\CHM{q}}
 }{
  \HALL
 }
 \Bigg(
  \frac{
   \EIGENx{k}{+}^2 \EIGEN{p}{} \EIGEN{q}{}
   -
   \EIGEN{p}{-} \EIGEN{q}{-}
  }{
   \EIGENx{k}{+} - \EIGENx{k}{-}
  }
  \nonumber\\&&
  +
  \frac{
   \EIGENx{k}{-}^2 \EIGEN{p}{} \EIGEN{q}{}
   -
   \EIGEN{p}{-} \EIGEN{q}{-}
  }{
   \EIGENx{k}{-} - \EIGENx{k}{+}
  }
 \Bigg)
 \EIGEN{p}{}\,
 \GEC{Z}{(\GEM{p}{};t)}\,
 \GEC{Z}{(\GEM{q}{};t)}\,
 \nonumber\\&&\hspace{-2em}
 =
 2 \pi \HEL_{k} \ABS{k}
 \mathop{
  \sum_{\GEM{p}{}} \sum_{\GEM{q}{}}
 }^{\vec{k}+\vec{p}+\vec{q}=\vec{0}}
 \tripleEuler{\big}{\CHM{k}}{\CHM{p}}{\CHM{q}}
 \EIGEN{p}{}^2\,
 \EIGEN{q}{}\,
 \GEC{Z}{(\GEM{p}{};t)}\,
 \GEC{Z}{(\GEM{q}{};t)}\,
\end{eqnarray}
Since
$
 \sum_{\GEM{q}{}}
 \EIGEN{q}{}
 \GEC{Z}{(\GEM{q}{};t)}
 \CHW{\phi}{(\CHM{q};\vec{x})}
 =
 \bm{b}
$
and
$
 \sum_{\GEM{p}{}}
 \EIGEN{p}{}^2
 \GEC{Z}{(\GEM{p}{};t)}
 \CHW{\phi}{(\CHM{p};\vec{x})}
 =
 \bm{u}-\HALL\bm{j}
,
$
the last line is the $\GEC{Z}{}$-representation 
of the evolution of the magnetic field (\ref{eq of mot b}).
The skew symmetry of 
$
 \tripleEuler{\big}{\CHM{k}}{\CHM{p}}{\CHM{q}}
$
between $\CHM{k}$ and $\CHM{q}$
leads to the identity
$$
 \mathop{
  \sum_{\GEM{k}{}} \sum_{\GEM{p}{}} \sum_{\GEM{q}{}}
 }^{\vec{k}+\vec{p}+\vec{q}=\vec{0}}
 \tripleEuler{\big}{\CHM{k}}{\CHM{p}}{\CHM{q}}
 \EIGEN{k}{}\,
 \EIGEN{p}{}^2\,
 \EIGEN{q}{}\,
 \GEC{Z}{(\GEM{k}{};t)}\,
 \GEC{Z}{(\GEM{p}{};t)}\,
 \GEC{Z}{(\GEM{q}{};t)}\,
 =
 0
$$
which yields the third constant of motion given by
\begin{eqnarray}
&&
 C_{3} =
 \sum_{\vec{k},\HEL_{k}}
 \frac{
  \left|\sum_{\POL_{k}}\EIGEN{k}{}\GEC{Z}{(\GEM{k}{};t)}\right|^2
 }{
  2\pi\HEL_{k}\ABS{k}
 }
 =
 \int_{\vec{x}\in M}
  \bm{b}\cdot\left((\nabla\times)^{-1}\bm{b}\right)
 {\rm{d}}^3\vec{x}
,
\end{eqnarray}
which is the magnetic helicity.

The evolution equation of 
$
 \EIGENx{k}{-} \GEC{Z}{(\GEMx{k}{+};t)}
 +
 \EIGENx{k}{+} \GEC{Z}{(\GEMx{k}{-};t)}
$
then
becomes
\begin{eqnarray}
&&\hspace{-3em}
 \frac{\partial}{\partial t}
 \left(
  \EIGENx{k}{-}
  \overline{\GEC{Z}{(\GEMx{k}{+};t)}}
  +
  \EIGENx{k}{+}
  \overline{\GEC{Z}{(\GEMx{k}{-};t)}}
 \right)
 \nonumber\\&&\hspace{-2em}
 =
 \mathop{
  \sum_{\GEM{p}{}} \sum_{\GEM{q}{}}
 }^{\vec{k}+\vec{p}+\vec{q}=\vec{0}}
 \frac{
  \tripleEuler{\big}{\CHM{k}}{\CHM{p}}{\CHM{q}}
 }{
  \HALL
 }
 \Bigg(
  \frac{
   -
   \EIGEN{p}{} \EIGEN{q}{}
   +
   \EIGENx{k}{-}^2 \EIGEN{p}{-} \EIGEN{q}{-}
  }{
   \EIGENx{k}{+} - \EIGENx{k}{-}
  }
  \nonumber\\&&
  +
  \frac{
   -
   \EIGEN{p}{} \EIGEN{q}{}
   +
   \EIGENx{k}{+}^2 \EIGEN{p}{-} \EIGEN{q}{-}
  }{
   \EIGENx{k}{-} - \EIGENx{k}{+}
  }
 \Bigg)
 \EIGEN{p}{}\,
 \GEC{Z}{(\GEM{p}{};t)}\,
 \GEC{Z}{(\GEM{q}{};t)}\,
\nonumber\\&&\hspace{-2em}
 =
 2 \pi \HEL_{k} \ABS{k}
 \mathop{
  \sum_{\GEM{p}{}} \sum_{\GEM{q}{}}
 }^{\vec{k}+\vec{p}+\vec{q}=\vec{0}}
 \tripleEuler{\big}{\CHM{k}}{\CHM{p}}{\CHM{q}}
 \EIGEN{q}{-}\,
 \GEC{Z}{(\GEM{p}{};t)}\,
 \GEC{Z}{(\GEM{q}{};t)}
.
\end{eqnarray}
Since
$
 \sum_{\GEM{p}{}}
  \GEC{Z}{(\GEM{p}{};t)}
  \CHW{\phi}{(\CHM{p};\vec{x})}
 =
 \bm{u}
$
and
$
 \sum_{\GEM{q}{}}
  \EIGEN{q}{-}
  \GEC{Z}{(\GEM{q}{};t)}
  \CHW{\phi}{(\CHM{q};\vec{x})}
 =
 \bm{b}+\HALL\nabla\times\bm{u}
,
$
the last line is the $\GEC{Z}{}$-representation of the evolution equation 
$
 \partial_{t}(\bm{b}+\HALL\nabla\times\bm{u})
 =
 \nabla\times(\bm{u}\times(\bm{b}+\HALL\nabla\times\bm{u}))
.
$
Skew symmetry of 
$
 \tripleEuler{\big}{\CHM{k}}{\CHM{p}}{\CHM{q}}
$
between $\CHM{k}$ and $\CHM{q}$
leads to the identity
$$
 \mathop{
  \sum_{\GEM{k}{}} \sum_{\GEM{p}{}} \sum_{\GEM{q}{}}
 }^{\vec{k}+\vec{p}+\vec{q}=\vec{0}}
 \tripleEuler{\big}{\CHM{k}}{\CHM{p}}{\CHM{q}}
 \EIGEN{k}{-}\,
 \EIGEN{q}{-}\,
 \GEC{Z}{(\GEM{k}{};t)}\,
 \GEC{Z}{(\GEM{p}{};t)}\,
 \GEC{Z}{(\GEM{q}{};t)}\,
 =
 0
$$
which yields the fourth constant of motion given by
\begin{eqnarray}
 C_{4}
 &=&
 \sum_{\vec{k},\HEL_{k}}
 \frac{
  \left| \sum_{\POL_{k}} \EIGEN{k}{-}\GEC{Z}{(\GEM{k}{};t)} \right|^2
 }{
  2\pi\HEL_{k}\ABS{k}
 }
\\
 &=&
 \int_{\vec{x}\in M}
  \big(\bm{b}+\HALL\nabla\times\bm{u}\big)
  \cdot
  \big((\nabla\times)^{-1}\bm{b}+\HALL\bm{u}\big)
 {\rm{d}}^3\vec{x}
,
\end{eqnarray}
which is the hybrid helicity.
The constant $C_{4}$ converges to $C_{3}$ in the limit $\HALL\to 0$.

Although these four constants of motion are derived naturally
from the symmetry of coefficients of the quadratic term
(\ref{rewrite the ((k||p||q))}),
$C_{2}$, $C_{3}$, and $C_{4}$ are not independent.
\renewcommand{\CalcNote}[1]{}%
Subtraction of $C_{3}$ from $C_{4}$ yields
\begin{eqnarray}
 C_{4} - C_{3}
 &=&
 \sum_{\vec{k},\HEL_{k}}
 \frac{
  \left| \sum_{\POL_{k}} \EIGEN{k}{-} \GEC{Z}{(\GEM{k}{};t)} \right|^2
 }{
  2\pi\HEL_{k}\ABS{k}
 }
 -
 \sum_{\vec{k},\HEL_{k}}
 \frac{
  \left| \sum_{\POL_{k}} \EIGEN{k}{} \GEC{Z}{(\GEM{k}{};t)} \right|^2
 }{
  2\pi\HEL_{k}\ABS{k}
 }
,
\CalcNote{
 \nonumber\\
 &=&
 \sum
 \frac{
  (\EIG_{\HEL}^{-})^2|\GEC{Z}{_{\HEL}^{+}}|^2
  +
  (\EIG_{\HEL}^{+})^2|\GEC{Z}{_{\HEL}^{-}}|^2
  +
  \EIG_{\HEL}^{+} \EIG_{\HEL}^{-} \left(
   \GEC{Z}{_{\HEL}^{+}} \overline{\GEC{Z}{_{\HEL}^{-}}}
   +
   \overline{\GEC{Z}{_{\HEL}^{+}}} \GEC{Z}{_{\HEL}^{-}}
  \right)
 }{
  2\pi\HEL\ABS{k}
 }
 \nonumber\\
 &&
 -
 \sum
 \frac{
  (\EIG_{\HEL}^{+})^2|\GEC{Z}{_{\HEL}^{+}}|^2
  +
  (\EIG_{\HEL}^{-})^2|\GEC{Z}{_{\HEL}^{-}}|^2
  +
  \EIG_{\HEL}^{+} \EIG_{\HEL}^{-} \left(
   \GEC{Z}{_{\HEL}^{+}} \overline{\GEC{Z}{_{\HEL}^{-}}}
   +
   \overline{\GEC{Z}{_{\HEL}^{+}}} \GEC{Z}{_{\HEL}^{-}}
  \right)
 }{
  2\pi\HEL\ABS{k}
 }
}
\nonumber\\
 &=&
 \HALL
 \sum
 \frac{
  \left( \EIGENx{k}{-}^2 - \EIGENx{k}{+}^2 \right)
  \left( |\GEC{Z}{(\GEMx{k}{+})}|^2 - |\GEC{Z}{(\GEMx{k}{-})}|^2 \right)
 }{
  \EIGENx{k}{+} + \EIGENx{k}{-}
 }
,
\nonumber\\
 &=&
 \HALL
 \sum
  \left( \EIGENx{k}{-} - \EIGENx{k}{+} \right)
  \left( |\GEC{Z}{(\GEMx{k}{+})}|^2 - |\GEC{Z}{(\GEMx{k}{-})}|^2 \right)
,
\nonumber\\
 &=&
 \HALL C_{2}
.
\end{eqnarray}
Thus, the constant $C_{2}$ is as follows:
\begin{eqnarray}
&&
 C_{2}=
 \int_{\vec{x}\in M} \left(
  2\bm{b}\cdot\bm{u}
  +
  \HALL
  \bm{u}\cdot(\nabla\times\bm{u})
 \right) {\rm{d}}^3\vec{x}
,
\label{modified cross helicity}
\end{eqnarray}
which converges to twice the cross helicity in the limit $\HALL \to 0$.
We refer to this quantity as the modified cross helicity hereafter.

\section{Discussion\label{sec:discussion}}

In this study an incompressible HMHD system was examined 
from a Lagrangian mechanical viewpoint.
%
%
The equation of motion of an incompressible HMHD medium
can be derived as a dynamical system on
a semidirect product of two diffeomorphism groups,
whose Lie algebra 
is
a function space of a pair of velocity and current fields.
The Riemannian metric is so defined that
the Lagrangian is given by the sum of the kinetic and magnetic energies.
%
%
Although the Lagrangian is same as that used in Hattori
to derive a standard MHD equation (Hattori 2011),
the difference between his configuration space and ours
leads to different dynamical systems.

To derive the evolution equation,
we followed the formulation given by Vizman ({Vizman 2001}).
%
%
When a Lie group
is extended by the semidirect product method
to obtain a specific dynamical system,
some vector space has been taken 
as a secondary set
and
the case of the semidirect product of two groups seems rare.
Note that,
the model we derived
includes the standard MHD equation in the limit $\HALL \to 0$
and
the Euler equation for an incompressible fluid 
for the $\bm{b}=\bm{0}$ case.

The Riemannian metric and commutator
can be defined 
on the function space of a pair of velocity and magnetic fields.
So
the function space constitutes a Lie algebra 
and
works as another configuration space of an HMHD system.
%
%
Thus,
the generalized {\Elsasser} variables
are shown to lead to remarkably simple expressions
for the components of the Riemannian metric, 
structure constants of the Lie algebra,
and
the equation of motion.

Comparisons between 
the HMHD system 
and 
the incompressible Euler equation system
should be made,
since this provides an insight into the nature of the conservation laws.
As had been shown by Arnold (Arnold 1966),
the Euler equation for an incompressible fluid 
$$
 \dot{\bm{u}}
 =
 \DivFree{\big}{
  \bm{u} \times
  (\nabla \times \bm{u})
 }
$$
is obtained as a dynamical system on $S$Diff($M$)
on which the Riemannian metric and Lie bracket are given by
$$
 \Braket{\big}{
  \bm{u}_{1}
 }{
  \bm{u}_{2}
 }_{e}
 :=
 \int_{\vec{\q}\in M}
  \bm{u}_{1} \cdot \bm{u}_{2} \,
 {\rm{d}}^3\vec{\q}
,\ \ \
 \LieBracket{\big}{\bm{u}_{1}}{\bm{u}_{2}}
 =
 \nabla \times ( \bm{u}_{1} \times \bm{u}_{2} )
,
$$
respectively.
Since the velocity field is divergence-free,
it can be expanded in CHW modes as
$
  \bm{u}(\vec{x},t)
  =
  \sum_{\CHM{k}}
    \CHC{u}{(\CHM{k};t)}\,
    \CHW{\phi}{(\CHM{k};\vec{x})}
,
$
and thus,
the Euler equation in CHW representation is given by
\begin{eqnarray}
\label{eq:euler eq}
  \frac{\partial}{\partial t}
    \overline{\CHC{u}{(\CHM{k};t)}}
  =
  \mathop{
    \sum_{\GEM{p}{}}
    \sum_{\GEM{q}{}}
  }^{\vec{k}+\vec{p}+\vec{q}=\vec{0}}
  \tripleEuler{\big}{\CHM{k}}{\CHM{p}}{\CHM{q}}
  \,\HEL_{q}\ABS{q}
  \,\CHC{u}{(\CHM{p};t)}
  \,\CHC{u}{(\CHM{q};t)}
.
\end{eqnarray}
Both equations
(\ref{basic equation in hatZ})
and
(\ref{eq:euler eq})
formally obey 
the geodesic equation
$\displaystyle
 \ddot{x}^i(t)
 =
 \Gamma_{jk}^{i}\,
 \dot{x}^j(t)\,
 \dot{x}^k(t)
.
$
This is a consequence of the fact that 
they are expressed
as a dynamical system on a specific Lie group.
Formal correspondence of mathematical objects
between these two systems is summarized 
in Table.\ref{Correspondence of mathematical objects}.
For the HMHD case,
the eigenvalue of the linear wave mode is given by 
a function of wavenumber other than a polynomial,
i.e.
its multiplication to eigenmode in wavenumber space
is an operation of a pseudodifferential operator.

\renewcommand{\GEM}[2]{\vec{#1},\HEL_{#1},{#2}\POL_{#1}}%
\renewcommand{\CHM}[1]{\vec{#1},{\HEL_{#1}}}%
\begin{table}[h]
\caption{%
\label{Correspondence of mathematical objects}
Correspondence of mathematical objects
between 
the Euler equation 
and 
the HMHD equation.
\vspace{1em}
}%
\hspace{-4em}
\begin{tabular}{c|cc}
\hline
  object & Euler eq. & Hall MHD system
\\\hline
  \begin{minipage}{6.5em}\centering
  basic variables
  \end{minipage}
  &
  ${\bm{u}}(\vec{x},t)$ &
  \begin{minipage}{15em}
  $\displaystyle
  \GEV{V}{(\vec{x},t)}=
  \left(\begin{array}{c}
    \bm{u}(\vec{x},t)
  \\
    -\HALL \bm{j}(\vec{x},t)
  \end{array}\right)
  $
  \\or\\
  $\displaystyle
  \GEV{Z}{(\vec{x},t)}=
  \left(\begin{array}{c}
    \bm{u}(\vec{x},t)
  \\
    \bm{b}(\vec{x},t)
  \end{array}\right)
  $
  \end{minipage}
\\&&
\\
  \begin{minipage}{6.5em}\centering
  intrinsic linear operator for normal modes
  \end{minipage}
  &
  $\nabla\times$ &
  $\displaystyle
    \left(\begin{array}{cc}
      O & I \\ I & -{\HALL}\nabla\times
    \end{array}\right)
  $
\\&&
\\
  eigenvalue & $2\pi\HEL\ABS{k}$ &
  $\displaystyle
    \EIGEN{k}{}
    =
    \HEL_{k}\left(
      \POL_{k}\sqrt{(\pi\HALL\ABS{k})^2+1}
      -\pi\HALL\ABS{k}
    \right)
  $
\\&&
\\
  \begin{minipage}{6.5em}\centering
  eigenfunction,
  normal mode
  \end{minipage}
  &
  \begin{minipage}{16em}
  complex helical wave:\\
  $\displaystyle
  \CHW{\phi}{(\CHM{k})}
  =
  \frac{\bm{e}_{\theta}(\vec{k})+i\HEL_{k}\bm{e}_{\phi}(\vec{k})}{\sqrt{2}}
  e^{2\pi i \vec{k}\cdot\vec{x}}
  $\\
  \end{minipage}
  &
  \begin{minipage}{17em}
  generalized {\Elsasser} variable:\\
  $\displaystyle
  \GEV{\Phi}{(\GEM{k}{})}
  =
  \left(\begin{array}{c}
    \CHW{\phi}{(\CHM{k};\vec{x})}
  \\
    \EIGEN{k}{}\,
    \CHW{\phi}{(\CHM{k};\vec{x})}
  \end{array}\right)
  $
  \end{minipage}
\\&&
\\
  mode indices &
  $\left\{\begin{array}{ll}
    \vec{k}: & \mbox{wavenumber,} \\
    \HEL: & \mbox{helicity,}
  \end{array}\right.$
  &
  $\left\{\begin{array}{ll}
    \vec{k}: & \mbox{wavenumber,} \\
    \HEL: & \mbox{helicity,} \\
    \POL: & \mbox{polarity,}
  \end{array}\right.$
\\&&
\\
  \begin{minipage}{6.5em}
  Riemannian metric tensor
  \end{minipage}
  &
  $\displaystyle
  \delta_{\HEL_{k}\HEL_{p}}
  \delta_{{k}_{x}{p}_{x}}
  \delta_{{k}_{y}{p}_{y}}
  \delta_{{k}_{z}{p}_{z}}
  $
  &
  \begin{minipage}{15em}
  $\displaystyle
  \big(1+(\EIG_{\HEL_{k}}^{{\POL}_{k}}({k}))^2\big)
  $\\
  $\displaystyle\qquad
  \times
  \delta_{\HEL_{k}\HEL_{p}}
  \delta_{{\POL}_{k}{\POL}_{p}}
  \delta_{{k}_{x}{p}_{x}}
  \delta_{{k}_{y}{p}_{y}}
  \delta_{{k}_{z}{p}_{z}}
  $
  \end{minipage}
\\&&
\\
  \begin{minipage}{5em}
  structure constant of Lie algebra\\
  \end{minipage}
  &
  $\displaystyle
    \tripleEuler{\big}{\CHM{k}}{\CHM{p}}{\CHM{q}}
    \HEL_{k}\ABS{k}
  $
  &
  $\displaystyle
    \frac{
      \tripleHMHD{\big}{\GEM{k}{}}{\GEM{p}{}}{\GEM{q}{}}
      \EIGEN{k}{-}
    }{
      1+\EIGEN{k}{}^2
    }
  $
\\
\hline
\end{tabular}
\end{table}

\renewcommand{\CHM}[1]{\hat{#1}}%

Formal correspondence of these two systems 
helps us to consider the mathematical foundation of 
the conservation laws derived in section 
\ref{constants of motion from Z-representation viewpoint}.
As was discussed,
the symmetry of the coefficient of $\GEC{Z}{}$-representation 
of an HMHD system derives \textit{four} conservation laws,
while two constants of motion, energy and helicity 
are known to be intrinsic in the Euler equation case
 ({Khesin and Chekanov 1989}).

Conservation of energy reflects the right-invariance 
of the Riemannian metric in each system.

Since
the Euler system counterpart of
the derivation process of the modified cross helicity
(\ref{eq:derivation C2 constant})
is
$$
  \mathop{
    \sum_{\CHM{k}{}}
    \sum_{\CHM{p}{}}
    \sum_{\CHM{q}{}}
  }^{\vec{k}+\vec{p}+\vec{q}=\vec{0}}
  \tripleEuler{}{\CHM{k}{}}{\CHM{p}{}}{\CHM{q}{}}
  \HEL_{k}\ABS{k}\,
  \HEL_{q}\ABS{q}\,
  \CHC{u}{(\CHM{k}{};t)}\,
  \CHC{u}{(\CHM{p}{};t)}\,
  \CHC{u}{(\CHM{q}{};t)}\,
  =
  0
,
$$
the modified cross helicity (\ref{modified cross helicity})
corresponds to the helicity in the Euler equation system,
which is given by
$
 H=\sum_{\CHM{k}{}}
 \HEL_{k}\ABS{k}\,
 |\CHC{u}{(\CHM{k}{};t)}|^2
 =
 \int\bm{u}\cdot(\nabla\times\bm{u})\,{\rm{d}}^3\vec{\q}
.
$
In the standard MHD limit ($\HALL \to 0$),
the modified cross helicity remains an $O(\HALL^0)$ quantity 
and converges to the cross helicity.

Thus,
energy and the modified cross helicity have their counterparts
in the Euler equation system.
Furthermore,
the derivation processes have their counterparts 
even in the freely spinning top case,
which is described by the Euler equation for a rigid body,
and thus,
these two constants appear to be 
more fundamental than the magnetic and hybrid helicities.

Magnetic and hybrid helicities are
expected to be intrinsic for HMHD systems,
since their derivation relies on 
the symmetry of the eigenvalue part of structure constant 
(\ref{rewrite the ((k||p||q))}).
This seems to reflect some unknown symmetry of an HMHD system
such that
the difference between the hybrid and magnetic helicities 
is proportional to the product of 
the Hall term coefficient and the modified cross helicity,
which vanishes in standard MHD limit.
Although
Noether's theory states that 
some conservation laws reflect 
the symmetries that a dynamical system intrinsically has,
investigation of the symmetry behind these two constants 
remains a subject for future work.


Despite the predominance of the studies from Hamiltonian mechanics viewpoint,
we took a Lagrangian mechanics approach,
i.e., variational formulation approach to the HMHD dynamics 
in the present study.
The advantage of this approach is that
it provides a differential geometrical framework 
to the predictability problem (Arnold 1966 section 11).
That is,
the second variation of the action 
is known to yield the Jacobi equation,
which provides information on the stability of paths
in the configuration space (Arnold 1989 appendix 1).
From this viewpoint,
attraction (stable) or repulsion (unstable) of infinitesimally close paths 
are determined in terms of the \textit{Riemannian curvature} 
associated with appropriate \textit{affine connection}
on which the time evolution is described as a \textit{geodesic}.
The value of the curvature is determined by
the snapshot pair of a (possibly time dependent) solution
and perturbation, i.e., the Jacobi field exerted on it
without solving the evolution equation or eigenvalue problem.
Thus,
the Lagrangian mechanical approach is the first step to 
the predictability problem  
and
we elucidated here the basic differential geometrical, 
Lie algebraic feature of configuration space of an HMHD system.
We are now working on our planned next article on this basis.

\vspace{1em}

\section*{Acknowledgment}
The author express his appreciation to Prof. Y. Hattori,
Prof. Y. Fukumoto and the anonymous referee 
for indication of some important references.
The author would also like to express his cordial gratitude to 
Professor Shigeyasu Kamiya for his continuous support and encouragement
during his career at OUS.
This work was performed under the auspices of 
the NIFS Collaboration Research Program (NIFS13KNSS044) 
and 
KAKENHI (Grant-in-Aid for Scientific Research(C)) 23540583.
The author would like to thank Enago for the English language review.

\renewcommand{\thesection}{Appendix \arabic{section}}
\setcounter{section}{0}
\section{Derivation of the value of Lie bracket in $\GEC{Z}$-representation
\label{derivation of the value of Lie bracket in Z-representation}}

\renewcommand{\EIGEN}[2]{\lambda_{\HEL_{#1}}^{{#2}\POL_{#1}}(\ABS{#1})}%
\renewcommand{\GEM}[2]{\tilde{#1}}%
\renewcommand{\CHM}[1]{\hat{#1}}%

Since CHW 
$
 \CHW{\phi}{(\CHM{k};\vec{x})}
$ 
is 
an eigenfunction of the curl operator 
and 
the corresponding eigenvalue is $2\pi\HEL_{k}\ABS{k}$,
the curl operation on CHW can be expressed in terms of 
eigenvalues of GEV as 
\begin{eqnarray}
 \renewcommand{\CHM}[1]{\vec{#1},{\HEL_{#1}}}
 \nabla\times \CHW{\phi}{(\CHM{k};\vec{x})}
 =
 -
 {\HALL}^{-1} (\EIGEN{k}{}+\EIGEN{k}{-})
 \CHW{\phi}{(\CHM{k};\vec{x})}
.
\label{eigenvalue of curl and GEV}
\end{eqnarray}

Substituting the CHW representation of 
the generalized {\Elsasser} variables (\ref{eigenfunction of L})
into the Lie bracket in $\GEV{Z}{}$-variable space 
(\ref{eq:Lie bracket in Z-space}),
taking the inner product defined by 
(\ref{eq:Riemannian metric in Z-space}),
and using the eigenvalue relations 
(\ref{sum of eigenvalues}),
(\ref{product of eigenvalues})
and
(\ref{eigenvalue of curl and GEV})
recursively,
we obtain the following formula:
\newcommand{\Q}{;\vec{x}}
\begin{eqnarray}
&&\hspace{-3em}
 \Braket{\Big}{
  \GEV{\Phi}{(\GEM{k}{})}
 }{
  \LieBrace{\Big}{
   \GEV{\Phi}{(\GEM{p}{})}
  }{
   \GEV{\Phi}{(\GEM{q}{})}
  }
 }_{Z}
 =
 \int_{\vec{x}\in M} {\rm{d}}^3\vec{x}\bigg[
  \CHW{\phi}{(\CHM{k}{\Q})} \cdot \nabla\times \Big(
   \CHW{\phi}{(\CHM{p}{\Q})} \times \CHW{\phi}{(\CHM{q}{\Q})}
  \Big)
  \nonumber\\&&\hspace{-2em}
  +
  \EIGEN{k}{} \CHW{\phi}{(\CHM{k}{\Q})} \cdot \Big(
   \CHW{\phi}{(\CHM{p}{\Q})} \times
   \big( \EIGEN{q}{} \nabla \times \CHW{\phi}{(\CHM{q}{\Q})} \big)
   +
   \big( \EIGEN{p}{} \nabla \times \CHW{\phi}{(\CHM{p}{\Q})} \big)
   \times \CHW{\phi}{(\CHM{q}{\Q})}
   \nonumber\\&&\hspace{-1em}
   -
   \HALL
   \big( \EIGEN{p}{} \nabla \times \CHW{\phi}{(\CHM{p}{\Q})} \big)
   \times
   \big( \EIGEN{q}{} \nabla \times \CHW{\phi}{(\CHM{q}{\Q})} \big)
  \Big)
 \bigg]
\nonumber
\\&&\hspace{-2em}
 =
 \tripleEuler{\big}{\CHM{k}}{\CHM{p}}{\CHM{q}}
 \bigg[
  -
  \frac{\EIGEN{k}{}+\EIGEN{k}{-}}{\HALL}
  +
  \EIGEN{k}{}\Big(
   -
   \EIGEN{q}{}
   \frac{\EIGEN{q}{}+\EIGEN{q}{-}}{\HALL}
   \nonumber\\&&\hspace{-1em}
   -
   \EIGEN{p}{}
   \frac{\EIGEN{p}{}+\EIGEN{p}{-}}{\HALL}
   -
   \HALL
   \EIGEN{p}{}
   \EIGEN{q}{}
   \frac{\EIGEN{p}{}+\EIGEN{p}{-}}{\HALL}
   \frac{\EIGEN{q}{}+\EIGEN{q}{-}}{\HALL}
  \Big)
 \bigg] 
\nonumber
\CalcNote{
\\&&\hspace{-2em}
 =
 \frac{
  \tripleEuler{\big}{\CHM{k}}{\CHM{p}}{\CHM{q}}
 }{
  \HALL
 }
 \bigg[
  -
  \big(\EIGEN{k}{}+\EIGEN{k}{-}\big)
  -
  \EIGEN{k}{}\big(\EIGEN{q}{}^2-1\big)
  -
  \EIGEN{k}{}\big(\EIGEN{p}{}^2-1\big)
  \nonumber\\&&
  -
  \EIGEN{k}{}\big(\EIGEN{p}{}^2-1\big)\big(\EIGEN{q}{}^2-1\big)
 \bigg] 
\nonumber
\\&&\hspace{-2em}
 =
 \frac{
  \tripleEuler{\big}{\CHM{k}}{\CHM{p}}{\CHM{q}}
 }{
  \HALL
 }
 \bigg(
  -
  \EIGEN{k}{-}
  -
  \EIGEN{k}{} \EIGEN{q}{}^2 \EIGEN{p}{}^2
 \bigg)
\nonumber
}
\\&&\hspace{-2em}
 =
 \frac{
  \tripleEuler{\big}{\CHM{k}}{\CHM{p}}{\CHM{q}}
 }{
  \HALL
 }
 \bigg(
  \EIGEN{k}{}^2 \EIGEN{p}{}^2 \EIGEN{q}{}^2 - 1
 \bigg)
 \EIGEN{k}{-}
,
\end{eqnarray}
where
$\tripleEuler{}{*}{*}{*}$
is defined by (\ref{def(k|p|q)}).

\section{Derivation of equation of motion in $\GEC{Z}{}$-representation
\label{derivation of equation of motion in Z-representation}}

Integrating by parts 
the first variation of action in $\GEV{Z}{}$-variable space 
(\ref{eq:first variation in Z-space}),
we obtain the extremal condition
\begin{eqnarray}
&&
 \int_{0}^{1}{\rm{d}}t \bigg[
  -
  \Braket{\Big}{
   \dot{\GEV{Z}{}}
  }{
   \GEV{\zeta}{}
  }_{Z}
  +
  \Braket{\Big}{
   {\GEV{Z}{}}
  }{
   \LieBrace{\Big}{
    \GEV{\zeta}{}
   }{
    \GEV{Z}{}
   }
  }_{Z}
 \bigg]
 = 0
\end{eqnarray}
Substituting (\ref{GEV expansion of Z}) 
and 
setting $\GEV{\zeta}{}=\GEV{\Phi}{(\GEM{k}{})}$,
we obtain the Euler-Lagrange equation
including the variation in the $\GEV{\Phi}{(\GEM{k}{})}$-direction
\begin{eqnarray}
&&\hspace{-3em}
 \Braket{\Big}{
  \sum_{{\GEM{q}{}}}
   \partial_{t}{\GEC{Z}{(\GEM{q}{})}}
   \GEV{\Phi}{(\GEM{q}{})}
 }{
  \GEV{\Phi}{(\GEM{k}{})}
 }_{Z}
 =
 \Braket{\Big}{
  \sum_{{\GEM{q}{}}}
   {\GEC{Z}{(\GEM{q}{})}}
   \GEV{\Phi}{(\GEM{q}{})}
 }{
  \LieBrace{\Big}{
   \GEV{\Phi}{(\GEM{k}{})}
  }{
   \sum_{{\GEM{p}{}}}
    {\GEC{Z}{(\GEM{p}{})}}
    \GEV{\Phi}{(\GEM{p}{})}
  }
 }_{Z}
.
\end{eqnarray}
Substituting 
(\ref{Riemannian metric in Z-coefficient}) 
and
(\ref{<k|{p,q}> in Z-space}),
we obtain the evolution equation of an HMHD system for 
$\GEC{Z}{}$-representation
\begin{eqnarray}
&&\hspace{-3em}
 (1+\EIGEN{k}{}^2)\partial_{t}\overline{\GEC{Z}{(\GEM{k}{})}}
 \nonumber\\&&
 =
 \sum_{{\GEM{q}{}}}
 \sum_{{\GEM{p}{}}}
 \frac{
  \tripleEuler{\big}{\CHM{q}}{\CHM{k}}{\CHM{p}}
 }{
  \HALL
 }
 \bigg(
  \EIGEN{q}{}^2 \EIGEN{k}{}^2 \EIGEN{p}{}^2 - 1
 \bigg)
 \EIGEN{q}{-}
 \GEC{Z}{(\GEM{p}{})}
 \GEC{Z}{(\GEM{q}{})}
.
\end{eqnarray}
Cycric symmetry of $\tripleEuler{}{*}{*}{*}$ 
leads to (\ref{basic equation in hatZ}).

\end{document}